\documentclass[showpacs,twocolumn,aps,prb]{revtex4-1}
\usepackage{amssymb}
\usepackage{amsmath}
\usepackage{graphicx}
\usepackage{subfigure}
\usepackage{color}
\usepackage{ulem}

\begin{document}
\title{Unidirectional Seebeck effect from Rashba spin-orbit coupling on the subsurface of  Ge(111)}

\author{Ge-Hui Zhu$^{1}$}

\author{Ying-Li Wu$^{1}$}

\author{Jia-Liang Wan$^{1}$}

\author{Xiao-Qin Yu$^{1}$}
\email{yuxiaoqin@hnu.edu.cn}

\affiliation{$^{1}$ School of Physics and Electronics, Hunan University, Changsha 410082, China.}

\begin{abstract}
A new nonlinear magnetothermal effect, namely unidirectional Seebeck effect(USE), has been recently reported in magnetic/nonmagnetic topological insulator (TI) heterostructure and is ascribed to the asymmetry magnon scattering. Here, we show a new mechanism to generate USE in the Rashba two dimensional electron gas (2DEG), in which the magnetism or magnetic order is completely absent. It's found that the USE has a quantum origin from the spin-momentum locking generated from Rashba spin-orbit coupling. We find that the USE exhibits a sine dependence on the orientation of the magnetic field with respect to the direction of the temperature gradient and is dominant from the states above the Lifshitz point. The USE in the semiconductor Ge(111) subsurface states has been theoretically and systematically investigated.
\end{abstract}

\maketitle
\section{Introduction}
\label{Introduction}
The spin-orbit coupling (SOC), an alternative to long-range magnetic order in semiconductors,
 has been playing a significant role in spintronics\cite{Wolf,Igor,Albert,David} and
gives birth to a tremendously active branch of spintronics: spin-orbitronics\cite{Manchon1,Wei,Manchon2,Trier}, which aims at discovering novel phenomena and functionalities stemming from SOC in a solid-state device.
Among SOCs, the Rashba spin-orbit coupling\cite{Rashba,Rashba2,Manchon3}, rooted in the inversion symmetry breaking and characterized with the spin splitting of electronic bands and spin-momentum locking in momentum, has attracted  great attention in spin-orbitronics and leads to a series of novel phenomena, such as bilinear magnetoresistance (BMR) \cite{Yang} , spin Hall effect (SHE)\cite{Manchon3,Anders}, valley Hall effect (VHE)\cite{Benjamin} and unidirectional magnetoresistance (UMR)\cite{Guillet}.



The unidirectional magnetoresistance \cite{Avci,Yasuda,Lv,He,He2,Ideue,Guillet}, referring to a change of the longitudinal magnetoresistance when reversing the polarity of current or the sign of the in-plane magnetization and manifesting, meanwhile, a linear dependence on both magnetic field and electric current itself, was first discovered in ferromagnet(FM)/normal metal (NM) bilayer\cite{Avci}, which stems from the modulation of the interface resistance by the SHE-induced spin accumulation. Subsequently, UMR has also been observed in topological insulator Bi$_2$Se$_3$\cite{He}, in two-dimensional electron gas at the SrTiO$_3$ (111) surface\cite{He2}, in polar semiconductor BiTeBr\cite{Ideue}, and in Rashba SOC induced Ge(111) subsurface states\cite{Guillet}. Owing to the absence of magnetic order in these cases, the effect has been related to the characteristic spin-momentum locking and has a distinct origin compared to the reported USE\cite{Avci,Yasuda,Lv} involving ferromagnetic layers. The UMR have attracted broad interest in nonlinear magnetoelectric and magnetothermal transport phenomena, such as BMR\cite{Dyrdal}, nonlinear planar Hall effect (NPE)\cite{He3}, nonlinear planar Nernst effect (NPNE)\cite{Yu,Zeng}, and unidirectional Seebeck effect (USE)\cite{Yu2}.

USE describes a nonlinear magnetothermal phenomenon\cite{Yu2}: The thermoelectric voltage $V$ generated from the Seebeck effect depends on the relative orientations of the in-plane magnetization with respect to the temperature gradient [Figs. \ref{figure1}(a) and (b)]; namely, when reversing the direction of temperature gradient with the fixed in-plane magnetic field or changing the sign of magnetic field with the fixed temperature gradient, the absolute value of generated voltage $|V|$ will be changed. The concept ``{unidirectional}" has appeared in the study of thermoelectric effects in the magnetic materials, for example, a unidirectional motion of the magnetic domain walls toward the hotter part of NiFe nanostrip \cite{Torrejon}, and a unidirectional spin wave propagation which gives thermal gradient and results in Seebeck voltage in a bilayer of a conducting polymer film and a magnetic insulator Y$_3$F$_5$O$_{12}$ (YIG)\cite{Pwang}. However, so far, the USE was only discovered in TI heterostructures composed of nonmagnetic TI(Bi$_{1-y}$Sb$_y$)$_{2}$Te$_3$(BST)\cite{Zhang1} and magnetic TI Cr$_x$(Bi$_{1-y}$Sb$_y$)$_{2-x}$Te$_3$ (CBST)\cite{Zhang2}, in which the involved Cr-doped surface layer effectively interacts with magnetism. The finite USE in the TI heterostructures originates from the asymmetric magnon scattering, namely asymmetric scattering of a conduction electron by a magnon due to the conservation of angular momentum \cite{Yu2}.

In this work, we will demonstrate that the USE could also exist in the semiconductor Ge (111) subsurface states, namely Rashba two-dimensional states (2DEG) localized in subsurface layers of Ge(111), in which the magnetism or magnetic order is completely absent. Our study show that USE is attributed to Rashba SOC, which generates spin-momentum locking in inside the subsurface states of semiconductor Ge (111). The concepts of USE is introduced in Sec.~\ref{CUSE}. The longitudinal nonlinear current $j^{(2)}_{x}$ as a second-order response to temperature gradient, giving rise to the USE, is derived for Ge (111) subsurface states in Sec.~\ref{TD}. The behaviors and mechanism of USE for Ge(111) subsurface states are discussed in Sec.~\ref{RAD}. Finally, we give a conclusion in Sec.~\ref{CON}.

\section{The concept of unidirectional Seebeck effect}
\label{CUSE}
 \begin{figure*}[htbp]
\centering
\flushleft
\includegraphics[width=1.0\linewidth,clip]{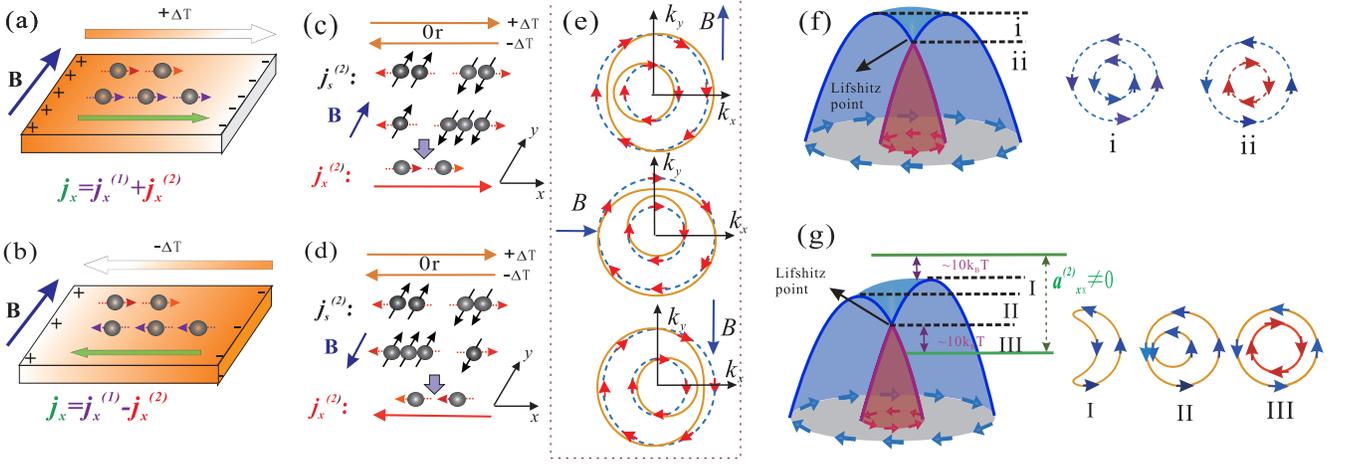}
\caption{Illustration of the concept of USE under (a) $+\Delta T$ and (b) $-\Delta T$ temperature difference. $j^{(1)}_{x}\propto (\nabla_{x}T)$ ($j^{(2)}_{x}\propto (\nabla_{x} T)^{2}$) are currents as the first-order (second-order) response to the temperature gradient, respectively.  (c),(d) Schematic illustration of the generation for the nonlinear current $j^{(2)}_{x}$ from the thermal driven nonlinear spin current $j^{(2)}_{s}$ by the temperature gradient $+\Delta T$ (or $-\Delta T$) when applying a magnetic field in (c) $+y$ axis and (d) $-y$ axis. The black and red arrows on the balls indicate the orientation of spin and the moving direction of carriers, respectively. The big long red (orange) arrow at the bottom (top) means the charge current flowing $j^{(2)}_{x}$ (the direction of temperature gradient), respectively.  (e) Illustration of the asymmetric distortion of Fermi contour of the upper band induced by $+ y$-direction (top), $+x$-direction (middle) and $-y$-direction (bottom) magnetic field $B$.  The Rashba energy band and its cross section in the absence of (f) and in the presence of (g) a magnetic field. The blue (red) dashed and yellow (red) solid curves represent the Fermi contours of the top (down) subsurface band without and with an external magnetic field, respectively.}
\label{figure1}
\end{figure*}
USE actually comes from the nonzero thermally driven nonlinear longitudinal current (NLC) $j^{(2)}_{x}\propto (\partial_{x} T)^{2}$ as a second-order response to temperature gradient [Figs. \ref{figure1}(a) and (b)]. When reversing the temperature gradient $\nabla_{x}T$, the thermal driven linear current $j^{(1)}_{x}\propto \partial_{x} T$ will change its direction correspondingly but with $j^{(2)}_{x}$ keeping unchanged. Consequently, the total current $j_{x}$ becomes different between the cases for the forward and backward temperature difference $\Delta T$, leading to a change of the generated voltage $|V|$ in the open circuit. In addition, although the direction of $j^{(2)}_{x}$ is independent on the direction of temperature gradient, it changes sign when reversing the magnetic field [Figs. \ref{figure1}(c) and (d)], namely $j^{(2)}_{x}(+\Delta T,+B)=j^{(2)}_{x}(-\Delta T,+B)=-j^{(2)}_{x}(+\Delta T,-B)$. Therefore, when reversing sign of the in-plane magnetic field with fixed temperature gradient, the generated voltage will also be  changed.

In this work, we find that the nonvanishing $j^{(2)}_{x}$ in Rashba 2DEG stems from the generation of the nonlinear longitudinal spin current (NLSC) $j^{(2)}_{s} $[Figs. \ref{figure1}(c) and (d)] as a second-order response to the temperature gradient, which could be converted into the longitudinal nonlinear current $j^{(2)}_{x}$ in the joint effect of in-plane magnetic field, Rashba SOC and particle-hole asymmetry. For example, when a magnetic field is applied in the $y$-direction, the Fermi contour will be distorted asymmetrically [top and bottom in Fig. \ref{figure1}(e)] owing to Rashba-SOC induced the spin-momentum locking and particle-hole asymmetry, causing an imbalance between the right- and left-moving electrons with opposite spin polarizations. As a result, a net second-order charge current $j^{(2)}_{x}$ will be generated from the NLSC [Figs. \ref{figure1}(c) and (d)].

Owing to the change of absolute value of voltage $|V|$ by reversing the direction of temperature gradient when fixing the in-plane magnetic field, one can easily find that the Seebeck coefficient $S$ is also noticeably different and temperature-gradient-direction-dependent in USE. Yu. \textit{et al.} have, phenomenologically, introduced a quantity $\Delta S$, which is the difference in Seebeck coefficients between the cases of the forward and backward temperature difference $\Delta T$ [Figs. \ref{figure1}(a) and (b)], to characterize the USE. The formula of $\Delta S$ has been derived in semiclassical framework of the electron dynamics as \cite{Yu}
\begin{equation}
\Delta S=\frac{2 \alpha^{\left(2\right)}_{xx} \Delta T}{\sigma_{xx}l}=\frac{2R_{xx}\alpha^{\left(2\right)}_{xx}w\Delta T}{l^{2}},
\label{S}
\end{equation}
where $\sigma_{xx}$ is the longitudinal conductivity, $l$ ($w$) refer to the length (width) of the sample, respectively; $R_{xx}$ denotes the longitudinal resistance, and  $\alpha^{\left(2\right)}_{xx}$ is the second-order response coefficient of the longitudinal nonlinear current $j^{(2)}_{x}$  to temperature gradient $\partial_{x}T$, namely  $j^{(2)}_{x}=\alpha^{\left(2\right)}_{xx}\left(\partial_{x}T\right)^{2}$. In the second equality, we have used $R_{xx}=l/(\sigma w)$ for the two-dimensional case. 
Therefore, when reversing the direction of temperature gradient, the difference of voltage $V_\text{USE}$ before and after reversing is determined by
\begin{equation}
V_\text{USE}=\Delta S \Delta T=2R_{xx}\alpha_{xx}^{(2)}w(\partial_{x}T)^{2}.
\end{equation}

Due to USE arising from the nonzero $j^{(2)}_{{x}}$, which is quantized by the coefficient $a^{(2)}_{xx}$, we will mainly focus on calculating and analyzing $j^{(2)}_{{x}}$ for Ge(111) subsurface states to reveal the origin and the behavior of the USE in this work. The signal of $V_\text{USE}$ will then be  estimated based on the calculated coefficient $a^{(2)}_{xx}$.

\section{The Thermally Driven nonlinear longitudinal current $j^{(2)}_{x}$ in Ge (111) subsurface states} \label{TD}
The Rashba SOC can generate spin-momentum locking inside the subsurface states of Ge [Fig. \ref{figure1}(f)], which have already been demonstrated in angle and spin-resolved photoemission spectroscopy experiments \cite{Ohtsubo1,Ohtsubo2,Aruga,Yaji}. The subsurface states of Ge are located above the maximum of the bulk valence band\cite{Guillet}. Therefore, the subsurface states could be segregated from the bulk states through modulating the Fermi energy into subsurface state by the gate voltage. As the Rashba band of subsurface states in the absence of magnetic field illustrated in Fig.~\ref{figure1}(f), the Fermi contour is made of two concentric rings with same spin helicities when the Fermi level locating in region i but opposite spin helicities when the Fermi level is below the Lifshitz point (region ii). In the presence of an in-plane magnetic field $\mathbf{B}=B(\cos\varphi,\sin\varphi)$ with the azimuth angle $\varphi$ measured from  the $x$-direction, the Rashba band is distorted [Fig.~\ref{figure1}(g)] and the electrons inside the subsurface state can be described by the following model Hamiltonian \cite{Guillet}:
\begin{equation}
H_{\mathbf{k}}\left(\textbf{B}\right)=-\frac{\hbar^{2}k^{2}}{2m^{\ast}}+\alpha_\text{R}\boldsymbol{\sigma}
\cdot(\textbf{k}\times\hat{\textbf{z}})+g\mu_{B}\boldsymbol{\sigma}\cdot\textbf{B},
\label{Hamiton}
\end{equation}
with $m^{\ast}$ being the effective mass of holes in the subsurface states, $\alpha_\text{R}$ denoting the Rashba spin-orbit interaction, $\sigma_{i=x,y,z}$ indicating the vector of Pauli matrices for spin, and $g$ and $\mu_{B}$ representing the land\'{e} factor and Bohr magneton, respectively. The first term generates the particle-hole asymmetry, the second term gives rise to a spin-momentum locking feature of energy band, and the third term denotes the Zeeman energy. The corresponding energy eigenvalue is
\begin{equation}
\epsilon_{n,\mathbf{k}}^{\text{B}}=
n \sqrt{(\alpha_\text{R}k_{y}+g\mu_{B}B_{x})^{2}+
(\alpha_\text{R}k_{x}-g\mu_{B}B_{y})^{2}}-
\frac{\hbar^{2}k^{2}}{2m^{\ast}},
\label{Enery}
\end{equation}
where the superscript ``{B}" represents magnetic field, and  $n=+1(-1)$ denotes the upper (lower) subsurface band, respectively.

The charge current in the $a$-direction is $j_{a}=-e\int[d\mathbf{k}]v_{a} f\left(\mathbf{r},\mathbf{k}\right)$, where the nonequilibrium distribution function response to the second order in temperature gradient can be expanded as $f=f_{0}+\delta f_{1}+\delta f_{2}$ with the term $\delta f_{n}$ vanishing as $(\partial T_{b}/ \partial r_{b})^{n}$. The first (second)-order nonequilibrium electron distribution $\delta f_{1}$ ($\delta f_{2}$) to  temperature gradient could be obtained via the iterative solution in the Boltzmann equation and have the following forms for a uniform and single-directional temperature gradient (see
Appendix \ref{APP-A-NEDF} for detailed discussion), respectively,
\begin{equation}
\begin{aligned}
\delta f_{1}\left(\mathbf{k}\right)&=\frac{\tau}{T\hbar}(\epsilon_\mathbf{k}-E_{f})\frac{\partial f_{0}}{\partial k_{b}}\partial_{b}T,\\
\delta f_{2}\left(\mathbf{k}\right)&=\tau^{2}\left[2\hbar v_{b}\frac{\partial f_{0}}{\partial k_{b}}+\left(\epsilon_\mathbf{k}-E_{f}\right)\frac{\partial ^{2}f_{0}}{\partial k_{b}^{2}}\right]\\
&\times\frac{\epsilon_\mathbf{k}-E_{f}}{\hbar^{2}T^{2}}(\partial_{b}T)^{2}.
\end{aligned}
\label{Core-f12}
\end{equation}
Hence, when applying the temperature gradient in the $x$-direction (i.e., $b=x$), the NLC $j^{(2)}_{x}=-\tau^{2}e\int[d\textbf{k}]v_{x}\delta f_{2}$ in the second-order temperature gradient is found to be
\begin{equation}
\begin{aligned}
j^{(2)}_{x}&=-\frac{\tau^{2}e}{\hbar^{2}T^{2}}\int[d\mathbf{k}]v_{x}(\epsilon_\mathbf{k}-E_{f})\left[2\hbar v_{x}\frac{\partial f_{0}}{\partial k_{x}}\right.\\
&\left.+\left(\epsilon_\mathbf{k}-E_{f}\right)\frac{\partial ^{2}f_{0}}{\partial k^{2}_{x}}\right]
(\partial_{x}T)^{2}.
\end{aligned}
\label{coef-jnl}
\end{equation}

 \begin{figure}[htbp]
\centering
\flushleft
\includegraphics[width=1.0\linewidth,clip]{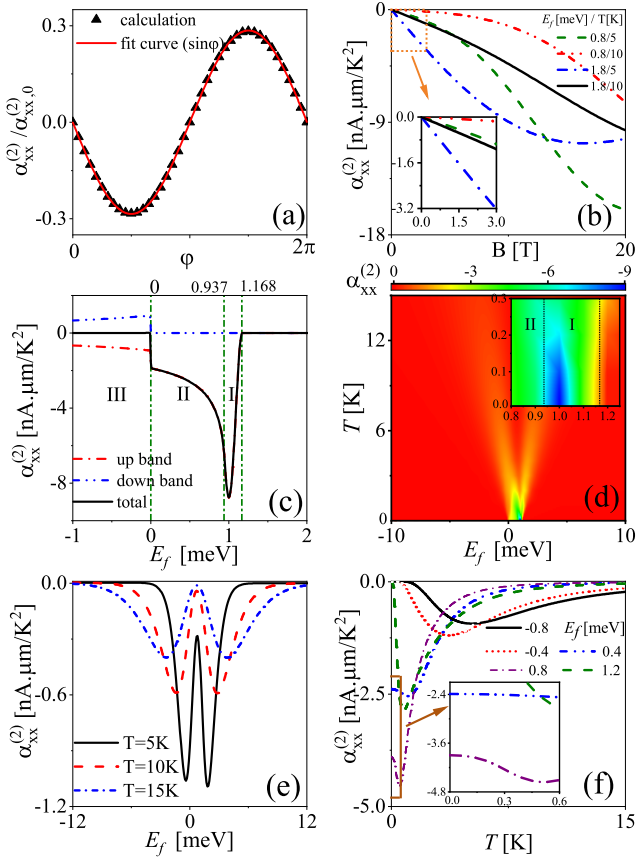}
\caption{(a) The in-plane magnetic field orientation (i.e., $\varphi$) dependence of normalized $a^{(2)}_{xx}/a^{(2)}_{xx,0}$, where $a^{(2)}_{xx,0}$ is $a^{(2)}_{xx}$ at $\varphi=\pi/2$.  (b) The coefficient $a^{(2)}_{xx}$ versus the magnetic field B for different Fermi energy $E_{f}$ and $T$. (c) $a^{(2)}_{xx}$ versus $E_{f}$ for different band at $T=10mK$. (d) The coefficient $a^{(2)}_{xx}$ as a function of Fermi energy $E_{f}$ and temperature gradient $T$.
(e) The coefficient $a^{(2)}_{xx}$ versus $E_{f}$ for different temperature $T$. (f) The coefficient $a^{(2)}_{xx}$ versus $T$ for several $E_{f}$.  $B=1 T$ is fixed in (c)-(f). Parameters used: $\alpha_\text{R}=-0.2~eV{\AA}$, $g=2$ and $\tau=2.88\times 10^{-11}s$.}
\label{figure2}
\end{figure}
Based on Eq.~(\ref{coef-jnl}), the coefficient $a_{xx}^{(2)}=j^{(2)}_{x}/(\partial_{x}T)^{2}$ can be determined. To numerically calculate $a_{xx}^{(2)}$, we use the following parameters for subsurface states inside Ge (111): $g=2$, the Rashba spin-orbit interaction $\alpha_\text{R}=-0.2~eV {\AA}$, \cite{Guillet,Ohtsubo1}  the effective mass $m=0.4 m_{e}$ \cite{Guillet} with $m_{e}$ being the electron mass, and the scattering relaxation time $\tau=2.28\times 10^{-11}s$ estimated by $\tau=\mu m /e$. The mobility $\mu$ of 2DEG in Ge can range from $3\times 10^{4}$ to $1.1\times 10^{6}~\text{cm}^{2}\text{V}^{-1}\text{s}^{-1}$.\cite{Myronov} $\mu=1\times10^{5}~\text{cm}^{2}\text{V}^{-1}\text{s}^{-1}$ is taken for an estimation.

\section{RESULTS AND DISCUSSION}
\label{RAD}
Figure \ref{figure2}(a) shows that $\alpha^{(2)}_{xx}$ exhibits $\sin\varphi$ dependence on the orientation of the in-plane magnetic field. This angle dependence is analogous to the USE originated from the magnon asymmetric scattering in which USE exhibits $\cos\phi$ dependence on the magnetization and the angle $\phi$ is measured from the $y$-direction\cite{Yu}. When the in-plane magnetic field is perpendicular to the temperature gradient (i.e., $\varphi=\pi/2,3\pi/2$), the quantity  $|\alpha^{(2)}_{xx}|$  will reach its maximum, giving rise to the largest value of USE. However, the USE will vanish (i.e., $\alpha^{(2)}_{xx}=0$) when the in-plane magnetic field is collinear with the temperature gradient.
\begin{figure*}[htbp]
\centering
\flushleft
\includegraphics[width=1.0\linewidth,clip]{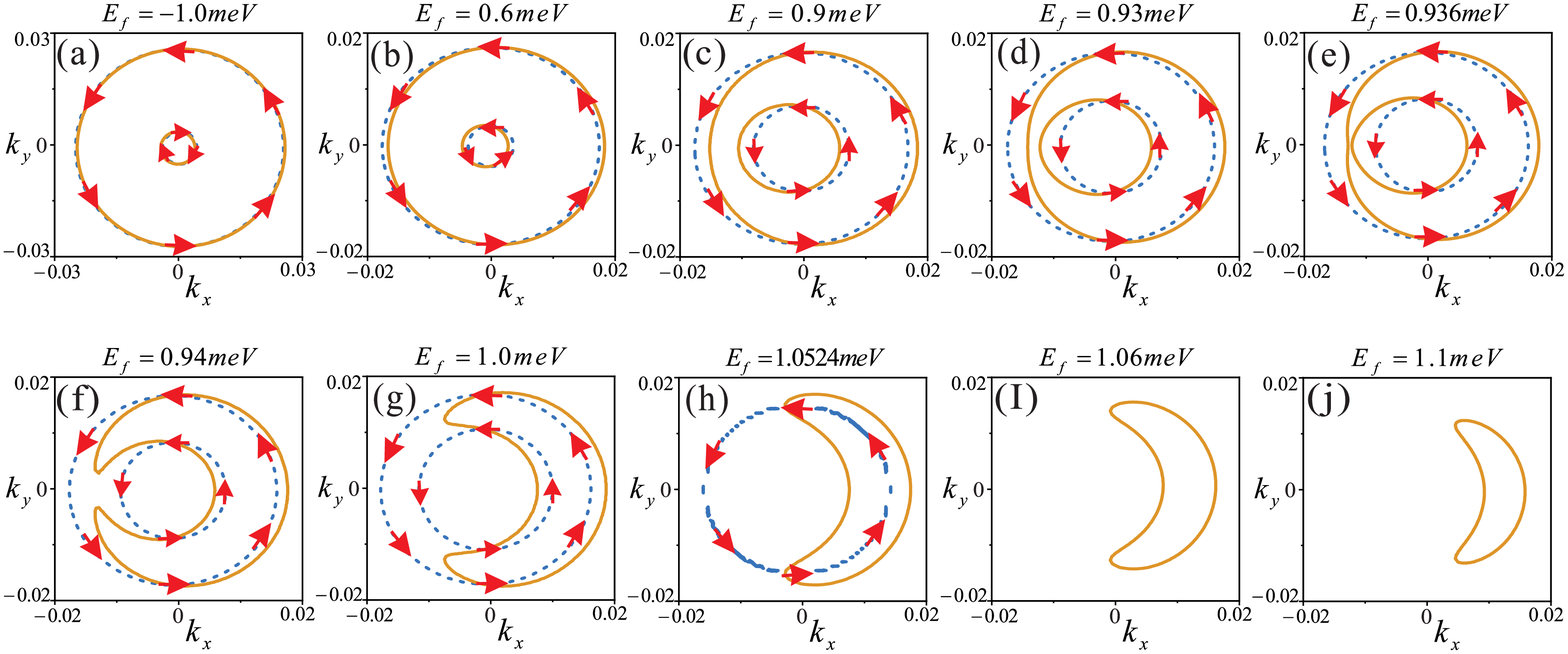}
\caption{Schematic illustration of the shift and  asymmetric distortion of the Fermi contour induced by the $y$-direction magnetic fields B for different Fermi energy $E_{f}$. The blue dashed (yellow solid) curves represent the Fermi contours of the subsurface band without (with) external magnetic field, respectively. The magnitude of magnetic field is fixed at $1T$. Momenta are measured in unit of ${\AA}^{-1}$ .
}
\label{figure3}
\end{figure*}

To understand the mechanism of USE and this $\sin\varphi$ dependence on the in-plane magnetic field, let's first analyze the thermally driven nonlinear longitudinal spin current $j^{(2)}_{s}$ and the influence of the temperature gradient on the nonequilbirium electron distribution in the absence of magnetic field. In fact, when applying temperature gradient to the Rashba-SOC induced subsurface, there is no linear spin current generated as a first-order response to the temperature gradient (see the following) and only $j^{(2)}_{s}$ will be generated.
The nonlinear spin current $[{j}^{(2)}_{s}]^{b}_{a}$ (where the subscript  ``{s}"  refers to spin) in the $a$-direction with spin pointing in the $b$-direction as a second-order response to temperature gradient is determined as
\begin{equation}
[{j}^{(2)}_{s}]^{b}_{a}=\frac{\hbar}{2}\int[d \textbf{k}]\langle \sigma^{b} \rangle v_{a}(\textbf{k})\delta f_{2}(\textbf{k}),
\end{equation}
where the average is carried out over the subsurface state of the upper (lower) band and can be replaced by $\langle\sigma^{b}\rangle=n h_{b}/|h|$ for $H_\mathbf{k}(\mathbf{B} = 0)$ with $\mathbf{h}=\alpha_\text{R}({\mathbf{k}}\times\hat{\mathbf{z}})$. In the absence of the magnetic field, the energy eigenvalue $\epsilon_{n,\mathbf{k}}^{0}$ is given as
\begin{equation}
\epsilon_{n,\mathbf{k}}^{0}=-\frac{\hbar^{2}k^{2}}{2m^{\ast}}+n|\alpha_\text{R}|k.
\label{eps1}
\end{equation}
According to Eq. (\ref{eps1}), one can easily observe that $\epsilon_{n,\mathbf{k}}^{0}$ are even functions with respect to $k_{x}$ and $k_{y}$, namely  $\epsilon_{n}^{0}\left(k_{x},k_{y}\right)=\epsilon_{n}^{0}\left(-k_{x},k_{y}\right)$ and $\epsilon_{n}^{0}\left(k_{x},k_{y}\right)=\epsilon_{n}^{0}\left(k_{x},-k_{y}\right)$, which are actually guaranteed by the joint constraints from the time-reversal symmetry and mirror symmetry of the Hamiltonian without the magnetic field [$H_\mathbf{k}(\mathbf{B} = 0)$ in Eq.(\ref{Hamiton})]. Therefore, when the magnetic field is absent, the energy and electron group $\mathbf{v}$  for the Rashba-SOC-induced subsurface states are even and odd in $\mathbf{k}$, respectively, which hints the first order of the temperature gradient $\delta f_{1}\left(\mathbf{k}\right)$ [Eq. (\ref{Core-f12})] is odd in $\mathbf{k}$ i.e., $\delta f_{1}\left(\mathbf{k}\right)=\delta f_{1}\left(-\mathbf{k}\right)$. Besides, $\langle\sigma^{b}\rangle$ is also odd in $\mathbf{k}$. Hence, from the parities, one could easily confirm $\frac{\hbar}{2}\int[d \textbf{k}]\langle \sigma^{b} \rangle v_{a}(\textbf{k})\delta f_{1}(\textbf{k})=0$, namely there is no linear spin current generated.

On the contrary, the second-order electron distribution function $\delta f_{2}\left(\mathbf{k}\right)$ [Eq.~(\ref{Core-f12})] is even in $\mathbf{k}$. In addition, electrons with $\mathbf{k}$ and $-\mathbf{k}$ carry opposite spins due to the Rashba-SOC-induced spin-momentum locking, namely, the spins of subsurface states are locked perpendicular to their momenta. As a result, the nonequilibrium surface states responding to the second order of temperature gradient with opposite momentum and opposite spins are equally populated in the absence of magnetic field, which leads to a nonzero $[{j}^{(2)}_{s}]^{a_{\perp}}_{a}$ with spin orientation in the $a_{\perp}$ direction owing to spin-momentum locking, hinting that the spin orientation of the thermally driven nonlinear spin longitudinal current points in $y$-direction, namely $[{j}^{(2)}_{s}]^{x}_{x}=0$ and $[{j}^{(2)}_{s}]^{y}_{x}\neq 0$.

When an in-plane magnetic field is applied, owing to the Rashba-SOC-induced spin-momentum locking and the electron-hole asymmetry, the Fermi energy contour will be shifted and distorted asymmetrically perpendicular to the magnetic field [Fig. \ref{figure1} (e)], causing the carriers with spin parallel to the magnetic field is no longer balanced with the opposite-moving ones with spin antiparallel to magnetic field. Thus, only the $y$ component of in-plane magnetic field $B_y$ will lead to the imbalance of carriers whose spins point along the $y$ axis and $-y$ axis, respectively. Consequently, a nonzero $j^{(2)}_{x}$, which gives rise to USE, as a second-order response to the applied temperature gradient will be generated from partially converting  $[{j}^{(2)}_{s}]^{y}_{x}$ with spin orientation along the $y$-direction into it and shows $\sin\varphi$ dependence on the in-plane magnetic field.

As expected, when applying the magnetic field along the $y$-direction, the coefficient $|\alpha^{(2)}_{xx}|$ increases monotonically with the increase of the magnitude of magnetic field $B$ [Fig.~\ref{figure2}(b)] since the shift and distortion of the Fermi energy will be strengthened with the enhancement of magnetic field. Besides, one can easily observe that $\alpha^{(2)}_{xx}$ manifests itself as a linear dependence on the magnetic field when $B<3T$. In fact, the $\sin\varphi$ dependence on the orientation and the linear dependence on the magnitude of the magnetic field could also be analytically analyzed when the magnetic field is small [$B<3T$] through the first-order expansion of Eq.~(\ref{coef-jnl}) in magnetic field [see details in Appendix \ref{TDNLC}].

In addition to the magnitude of magnetic field, the strength of shift and distortion of Fermi contour are also affected by the Fermi level.  Figure \ref{figure3} illustrates the distortion of Fermi contour induced by an in-plane magnetic field with $\mathbf{B}=(0,1T)$  for subsurface states of Ge at different Fermi level. When $B_{x}=0$ and $B_{y}=1T$, the corresponding energy ranges for the divided regions [Fig.~\ref{figure1}(g)] are given as follows: when the energy is below the Lifshitz point (namely $E_{f}<0$), it's classified into region $\text{III}$ in which both the upper and lower band exist and they show opposite spin helicities; when the energy is in the range of [0,~0.937 $meV$], the Fermi level lies in region $\text{II}$ in which two distorted rings from the upper band show same spin helicities; when the Fermi level is in the energy scale from $0.937~meV$ to $1.168~meV$, it belongs to region $\text{I}$ in which only the upper band exits.

One could easily observe that the shift and distortion are enhanced with increase of  $E_{f}$ in region II ($0<E_{f}<0.937~meV$) [Fig.~\ref{figure3}], leading to the monotonically increase of $|a_{xx}^{(2)}|$ [Fig.~\ref{figure2}(c)]. With further increasing the Fermi level into the region I ($0.937<E_{f}<1.168~meV$), the Fermi contour first changes from the two connected distorted rings into a crescent moon leading to the enhancement of imbalance between right- and left-moving electrons with opposite spin polarizations , and then the crescent moon gradually shrinks and finally disappears. As a result, the $|a_{xx}^{(2)}|$ would first increase owing to the enhancement of carrier imbalance and then gradually decrease due to the shrinking crescent moon. Besides, once decreasing $E_{f}$ into region III, the signal of $a_{xx}^{(2)}$ becomes zero or very weak for the very low temperature since the converted nonlinear charge current from spin nonlinear current for each subband has opposite signs with respect to each other owing the opposite spin helicities for the upper and lower band and hence, contribution of the two subbands to $a^{(2)}_{xx}$ cancel or partially cancel [Fig.~\ref{figure2}(c)].

Therefore, owing to those distortions of Fermi contour changing with Fermi level, the nonvanishing $|a^{(2)}_{xx}|$ exits in regions I and II, and a peak feature manifests itself in region I [Fig. \ref{figure2}(c)]. However, this behavior of $|a^{(2)}_{xx}|$ vs $E_{f}$ is only true for very low temperature, for example $T=10m\text{K}$ [Fig. \ref{figure2}(c)], in which the temperature broadening effect of Fermi distribution is negligible. 
As shown in Figs. \ref{figure2}(d) and (e), with the temperature increasing, the two peak feature ($|a^{(2)}_{xx}|$ vs $E_{f}$) will appear instead of a peak feature possibly owing to the combination of the temperature broadening effect of Fermi distribution and the distortion of Fermi contour. The magnitude of peaks decrease with increasing of temperature and positions of right (left) peak shifted towards to the higher (lower) Fermi level. Besides, the nonvanishing $a^{(2)}_{xx}$ can also exist when the Fermi level goes beyond the region I and II within $10k_{B}T$ [Fig.~{\ref{figure1}}(g)]. The temperature dependence of $a^{(2)}_{xx}$ at different Fermi energy levels is shown in Fig.~{\ref{figure2}}(f). As expected, when the Fermi energy goes beyond regions I and II, $a_{xx}^{(2)}$ tends to zero when $T$ approaches zero owing to the opposite spin helicities of two subbands in regions III or no charge carriers above region I. However, when modulating the Fermi level into regions I and II, $a_{xx}^{(2)}$ tends to be a constant when the temperature approaches zero. The largest values of $|a_{xx}^{(2)}|$ appear in regions I and {\color{blue}{II}} at low temperature [inset of Fig.~\ref{figure2}(d)].

To numerically estimate the signal of USE [$V_\text{USE}=2R_{xx}\alpha_{xx}^{(2)}w(\partial_{x}T)^{2}$] stemmed from Rashba SOC in Ge (111) subsurface, we take $\alpha^{(2)}_{xx}=3.2~\text{nA}\mu \text{m}/\text{K}^{2}$ [Fig. \ref{figure2}(b)] for $T=5\text{K}$ and $B=3\text{T}$. The longitudinal resistance $R_{xx}$ of Ge(111) subsurface for $T=5~K$ is in range of $1$-$3.65~k\Omega$ when the Fermi energy is located in region I and II [Fig.~\ref{figure4}]. We use $R_{xx}=3.65k~\Omega$ for an estimation. In experiment, the temperature gradient can already reach $1.5~\text{K}\mu m^{-1}$.\cite{Xu} Therefore, the difference of voltage $V_\text{USE}$ generated from USE can reach $1.57~mV$ with $w=30~\mu m$, which is an order of magnitude smaller than that reported in TI heterostructure Cr$_{x}$(Bi$_{1-y}$Sb$_y$)$_{2-x}$Te$_{3}$/(Bi$_{1-y}$Sb$_{y})_{2}$Te$_{3}$ but still within  measurable range in experiment\cite{Uchidak}.


\section{Conclusion}
\label{CON}
In summary, we have shown that a nonlinear unidirectional Seebeck effect can emerge in Rashba 2DEG in the complete absence of magnetism and magnetic order. It's found that the USE originates from the conversion of a nonlinear longitudinal spin current to a charge current due to the band bending induced by the joint result of spin-momentum locking generated from Rashba SOC, hole-electron asymmetry and the in-plane magnetic field. The USE strongly depends on the orientation of the in-plane magnetic field and shows $\sin\varphi$ dependence on the magnetic field. The unidirectional Seebeck effect in the semiconductor Ge(111) subsurface states have been theoretically and systematically investigated. The difference of the voltage $V_\text{USE}$ quantifying the USE in Ge(111) is an order of magnitude smaller than that reported in TI heterostructure Cr$_{x}$(Bi$_{1-y}$Sb$_y$)$_{2-x}$Te$_{3}$/(Bi$_{1-y}$Sb$_{y})_{2}$Te$_{3}$ but is still within  measurable range in experiment. A larger USE in Rashba 2DEG might be expected in the materials with strong Rashba spin-orbit coupling, such as the surface of the topological insulator Bi$_2$Se$_3$\cite{King} and the surfaces of Bismuth Tellurohalides BiTeCl\cite{Eremeev}. Besides, the analysis of the coefficient $a^{(2)}_{xx}$ dependence on the Fermi energy at low temperature suggests that the USE is predominantly from the states above the Lifshitz point, namely regions I and II in Fig.~\ref{figure1}(g).

\section*{ACKNOWLEDGMENTS}
The work is supported by the Fundamental Research Funds for the Central Universities and NSFC (Grant No. 12004107).

\appendix

\setcounter{equation}{0}
\setcounter{figure}{0}
\setcounter{table}{0}
\makeatletter
\renewcommand{\theequation}{A\arabic{equation}}
\renewcommand{\thefigure}{A\arabic{figure}}
\renewcommand{\thetable}{A\arabic{table}}

\bigskip
\bigskip

\noindent
\section{Derivation of the non-equilibrium distribution function that responds to the temperature gradient}
\label{APP-A-NEDF}
In the absence of the external electric field, the Boltzmann equation for the electron distribution within the relaxation time approximation is
\begin{equation}
f-f_{0}=-\tau \frac{\partial f}{\partial r_{a}}\cdot v_{a},
\label{App-A-1}
\end{equation}
with $f_{0}=1/(\textrm{exp}[\frac{\epsilon_{\mathbf{k}}-E_{f}}{k_{B}T}]+1)$ indicating the equilibrium Fermi distribution, $\tau$ denoting the relaxation time, and $r_{a}$ and $v_{a}$ representing the $a$ components of coordinate position and velocity of electrons, respectively.
The local distribution function as response up to the second order in temperature gradient $\nabla T$ can be expressed as:
\begin{equation}
\begin{aligned}
f(\textbf{r},\textbf{k})&= f_{0}(\textbf{r},\textbf{k})+A_{a}\frac{\partial T}{\partial r_{a}}+B_{ab}\frac{\partial T}{\partial r_{a}}\frac{\partial T}{\partial r_{b}}+O[(\partial_{a}T)^{3}]\\
&\approx f_{0}(\textbf{r},\textbf{k})+\delta f_{1}(\partial_{a}T)+\delta f_{2}(\partial_{a}T\partial_{b}T),
\end{aligned}
\label{App-dis}
\end{equation}
with $a,b=x\, , \text{or} \, y$. In the second line of Eq.~(\ref{App-dis}), for convenience, we have made $\partial_{a}=\partial/\partial r_{a}$ and
\begin{equation}
\begin{aligned}
\delta f_{1}(\partial_{a}T)&=A_{a}\partial_{a}T,\\
\delta f_{2}(\partial_{a}T\partial_{b}T)&=B_{ab}\partial_{a}T\partial_{b}T.
\end{aligned}
\end{equation}
Owing to the local equilibrium distribution function $f_{0}(\textbf{r},\textbf{k})$ fixed itself by the temperature at $\textbf{r}$,\cite{Ziman} one has
\begin{equation}
\begin{aligned}
\frac{\partial f_{0}}{\partial r_{a}}=\frac{\partial f_{0}}{\partial T}\frac{\partial T}{\partial r_{a}}=-\frac{(\epsilon_\mathbf{k}-E_{f})}{T}\frac{\partial f_{0}}{\partial \epsilon_{\textbf{k}}}\partial_{a}T.
\end{aligned}
\label{App-A-T}
\end{equation}
Combining Eq.~(\ref{App-A-T}) with the equality $\frac{\partial \epsilon_{\textbf{k}}}{\partial \textbf{k}}=\hbar \textbf{v}$, we can further  transform $\frac{\partial f_{0}}{\partial T}$ into $\frac{\partial f_{0}}{\partial \textbf{k}}$ through differential treatment
\begin{equation}
\begin{aligned}
\frac{\partial f_{0}}{\partial \textbf{k}}=\frac{\partial f_{0}}{\partial \epsilon_{\textbf{k}}}\cdot \frac{\partial \epsilon_{\textbf{k}}}{\partial \textbf{k}}=-\frac{\hbar \textbf{v}T}{(\epsilon_{\textbf{k}}-E_{f})}\frac{\partial f_{0}}{\partial T},
\end{aligned}
\end{equation}
giving the following identities
\begin{equation}
\begin{aligned}
\frac{\partial{f_{0}}}{\partial T} \cdot v_{a}&=-\frac{\epsilon_{k}-E_{f}}{\hbar T}\frac{\partial f_{0}}{\partial k_{a}},\\
\frac{\partial^{2}f_{0}}{\partial^{2}T}v_{a}v_{b}&=2\frac{\epsilon_{k}-E_{f}}{\hbar T^{2}}\frac{\partial f_{0}}{\partial k_{a}}v_{b}+\left(\frac{\epsilon_{k}-E_{f}}{\hbar T}\right)^{2}\frac{\partial^{2} f_{0}}{\partial k_{a}\partial k_{b}}.
\end{aligned}
\label{App-A-iden}
\end{equation}
Substituting the formula of $f$ in Eq.~(\ref{App-dis}) into Eq.~(\ref{App-A-1}) and comparing the expansion coefficients for the first-order $\partial_{a}T$, one can easily find that
\begin{equation}
\delta f_{1}(\partial_{a}T)=-\tau\frac{\partial f_{0}}{\partial r_{a}}\cdot v_{a}=-\tau \frac{\partial f_{0}}{\partial T}\partial_{a}T\cdot v_{a}.
\label{App-A-f11}
\end{equation}
Through iteration, we can have
\begin{equation}
\begin{aligned}
\delta f_{2}(\partial_{a}T\partial_{b}T)&=-\tau\frac{\partial}{\partial r_{a}}(\delta f_{1}) v_{a}\\
&=\tau^{2}(\frac{\partial^{2}f_{0}}{\partial^{2}T}\partial_{a}T\partial_{b}T+\frac{\partial f_{0}}{\partial T}\partial_{ab}T)v_{a}v_{b}.
\end{aligned}
\label{App-A-f12}
\end{equation}
Assuming the uniform  temperature gradient in the system, namely $\partial_{ab}T =0$, and accompanying with Eq.~(\ref{App-A-iden}), the nonequilibrium distribution function $\delta f_{1}\left(\mathbf{k}\right)$ [Eq.(\ref{App-A-f11})] and  $\delta f_{2}\left(\mathbf{k}\right)$ 
[Eq.(\ref{App-A-f12})] as the first-order and  second-order response to the temperature gradient are found to be, respectively,
\begin{equation}
\begin{aligned}
\delta f_{1}\left(\mathbf{k}\right)&=\frac{\tau}{T\hbar}(\epsilon_\mathbf{k}-E_{f})\frac{\partial f_{0}}{\partial k_{a}}\partial_{a}T,\\
\delta f_{2}\left(\mathbf{k}\right)&=\tau^{2}\left[2\hbar v_{b}\frac{\partial f_{0}}{\partial k_{a}}+\left(\epsilon_\mathbf{k}-E_{f}\right)\frac{\partial ^{2}f_{0}}{\partial k_{a}\partial k_{b}}\right]\\
&\times\frac{\epsilon_\mathbf{k}-E_{f}}{\hbar^{2}T^{2}}\partial_{a}T\partial_{b}T.
\end{aligned}
\label{App-A-F12F}
\end{equation}
It should be pointed out that if the temperature gradient is applied in a single direction, one can have $a=b$ in Eq. (\ref{App-A-F12F}).

\makeatletter
\renewcommand{\theequation}{B\arabic{equation}}
\renewcommand{\thefigure}{B\arabic{figure}}
\renewcommand{\thetable}{B\arabic{table}}
\section{The thermally driven nonlinear longitudinal current to the first-order of magnetic field}
\label{TDNLC}

With the formula for $\delta f_{2}(\mathbf{k})$ determined in Eq.~(\ref{App-A-F12F}), the $a$-component of nonlinear current $j^{(2)}_{a}=-\tau^{2}e\int[d\textbf{k}]v_{a}\delta f_{2}$ in the second-order temperature gradient is found to be
\begin{equation}
\begin{aligned}
j^{(2)}_{a}&=-\frac{\tau^{2}e}{\hbar^{2}T^{2}}\int[d\mathbf{k}]v_{a}(\epsilon_\mathbf{k}-E_{f})\left[2\hbar v_{b}\frac{\partial f_{0}}{\partial k_{a}}\right.\\
&\left.+\left(\epsilon_\mathbf{k}-E_{f}\right)\frac{\partial ^{2}f_{0}}{\partial k_{a}\partial k_{b}}\right]\partial_{a}T\partial_{b}T.
\label{App-B-j2}
\end{aligned}
\end{equation}
To first-order approximation of the magnetic field, the thermally driven nonlinear current $j^{(2)}_{a}$ can be expanded as
\begin{equation}
j^{(2)}_{a}=\sum_{bc}W_{abc}\partial_{b}T\partial_{c}T+\sum_{bcd}U_{abcd}\partial_{b}T\partial_{c}
TB_{d},
\end{equation}
where the nonlinear response functions $W_{abc}$ and $U_{abcd}$ are
\begin{widetext}
\begin{equation}
\begin{aligned}
W_{abc}&=\frac{-\tau^{2}e}{T^{2}\hbar^{2}}\int[d\mathbf{k}]v_{a}\left[\left(
\epsilon^{0}_\mathbf{k}-E_{f}\right)\hbar v_{b}\frac{\partial f_{0}}{\partial k_{c}}
+\left({\epsilon^{0}_\mathbf{k}-E_{f}}\right)^{2}\frac{\partial^{2}f_{0}}{\partial k_{b}\partial k_{c}}
\right],\\
U_{abcd}=&\frac{-\tau^{2}egu_{B}}{T^{2}\hbar^{2}}\int[d\mathbf{k}]
\left[\left({\epsilon^{0}_\mathbf{k}-E_{f}}\right)^{2}\left(\frac{\partial^{2} f_{0}}{\partial k_{b}\partial k_{c}}\frac{\partial v_{a}}{\partial h_{d}}+\frac{v_{a}\partial^{3} f_{0}}{\partial k_{b}\partial k_{c}\partial h_{d}}\right)+2\frac{\partial \epsilon^{0}_\mathbf{k}}{\partial h_{d}}\hbar v_{a}v_{b}\frac{\partial f_{0}}{\partial k_{c}}+2\left(\epsilon^{0}_\mathbf{k}-E_{f}\right)\frac{\partial \epsilon^{0}_\mathbf{k}}{\partial h_{d}}v_{a}\frac{\partial^{2}f_{0}}{\partial k_{b}\partial k_{c}}\right.\\
&\left.+2\left({\epsilon^{0}_\mathbf{k}-E_{f}}\right)\hbar\left(\frac{\partial^{2} f_{0}}{\partial k_{c}\partial h_{d}}v_{a}v_{b}+\frac{\partial f_{0}}{\partial k_{c}}\frac{\partial v_{a}}{\partial h_{d}}v_{b}+\frac{\partial f_{0}}{\partial k_{c}}\frac{\partial v_{b}}{\partial h_{d}}v_{a}\right)\right],
\end{aligned}
\label{App-B-RF1}
\end{equation}
\end{widetext}
where $\epsilon^{0}_\mathbf{k}$ indicates the energy in absence of magnetic field and is given in Eq.~(\ref{eps1}). For the simplicity, the band index $n$ is ignored. To obtain Eq. (\ref{App-B-RF1}), one need to use the relation $\frac{\partial F(\epsilon_\mathbf{n,k}^{M})}{\partial B_{d}}=g\mu_{B}\frac{\partial F(\epsilon_\mathbf{n,k}^{M})}{\partial h_{d}}$ with $\mathbf{h}=\alpha_\text{R}
(\textbf{k}\times\hat{\textbf{z}})$.
Through exploiting the parities (namely even/ odd function with respect to $k_{x}$ and $k_{y}$) of the integrand in Eq.(\ref{App-B-RF1}), the following tensor elements are found to be zero
\begin{equation}
\begin{aligned}
W_{abc}&=0, \quad a,b,c=x,y\\
U_{xyyx}&=U_{xyxy}=U_{xxyy}=U_{xxxx}=0,\\
U_{yxxy}&=U_{yxyx}=U_{yyxx}=0,\\
\end{aligned}
\label{coefficients}
\end{equation}
which hints that, when applying a temperature gradient $\partial_{x}T$ along the $x$-direction (i.e., $b=c=x$) and an in-plane magnetic field $\mathbf{B}=B\left(\cos\varphi,\sin\varphi\right)$, the nonlinear longitudinal current $j^{(2)}_{x}$ flowing along the $x$-direction (i.e., $a=x$) as the second-order response to temperature gradient  is found to be
\begin{equation}
\begin{aligned}
j^{(2)}_{x}&=\left[W_{xxx}+\left(U_{xxxx}\cos\varphi+U_{xxxy}\sin\varphi\right)B\right]
(\partial_{x}T)^{2}\\
&=U_{xxxy}\sin\varphi B\left(\partial_{x}T\right)^{2},
\end{aligned}
\label{App-B-NL-CU}
\end{equation}
where the nonlinear planar Seebeck coefficient $U_{xxxy}$ is determined as
\begin{equation}
\begin{aligned}
U_{xxxy}&=\frac{-\tau^{2}eg\mu_{B}}{T^{2}}\int [d\textbf{k}]\Big[\frac{\partial f_{0}}{\partial \epsilon^{0}_\mathbf{k}}2\Upsilon_{1}+\frac{2(\epsilon^{0}_\mathbf{k}-E_{f})}{\hbar}(\frac{\partial f_{0}}{\partial \epsilon^{0}_\mathbf{k}}\Upsilon_{2}\\
&+\frac{\partial^{2} f_{0}}{\partial (\epsilon^{0}_\mathbf{k})^{2}}2\hbar \Upsilon_{1})+(\frac{\epsilon^{0}_\mathbf{k}-E_{f}}{\hbar})^{2}(\frac{\partial f_{0}}{\partial \epsilon^{0}_\mathbf{k}}\hbar \Upsilon_{3}+\frac{\partial^{2}f_{0}}{\partial (\epsilon^{0}_\mathbf{k})^{2}}\hbar \Upsilon_{2}\\
&+\frac{\partial^{3}f_{0}}{\partial (\epsilon^{0}_\mathbf{k})^{3}}\hbar^{2}\Upsilon_{1})\Big],
\end{aligned}
\end{equation}
with
\begin{equation}
\begin{aligned}
\Upsilon_{1}&=v_{x}^{3} v_{h_{y}}\\
\Upsilon_{2}&=3\hbar v_{x}^{2}v_{xh_{y}}+\hbar v_{x}v_{xx} v_{h_{y}}\\
\Upsilon_{3}&= v_{xx}v_{xh_{y}}+v_{x} v_{xxh_{y}}
\end{aligned}
\end{equation}
where $v_{x}=\partial \epsilon^{0}_\mathbf{k}/(\hbar \partial k_{x})$ represents the $x$ component of electron velocity in the absence of a magnetic field. Here, for simplicity, the coefficients $v_{xx}=\partial v_{x}/\partial k_{x}$, $v_{h_{y}}=\partial \epsilon^{0}_\mathbf{k}/(\hbar \partial h_{y})$, $v_{xh_{y}}=\partial v_{x}/\partial h_{y}$, and $v_{xxh_{y}}=\partial v_{xx}/\partial h_{y}$ have been introduced. 

Combining the definition of $\alpha^{\left(2\right)}_{xx}$ with Eq. (\ref{App-B-NL-CU}), one would easily verify that
\begin{equation}
\alpha^{\left(2\right)}_{xx}=U_{xxxy}B\sin\varphi ,
\end{equation}
showing a $\sin\varphi$ dependence of $\alpha^{\left(2\right)}_{xx}$ on the orientation of the magnetic field and the linear dependence on the magnitude of the magnetic field.

\makeatletter
\renewcommand{\theequation}{C\arabic{equation}}
\renewcommand{\thefigure}{C\arabic{figure}}
\renewcommand{\thetable}{C\arabic{table}}

\section{The longitudinal Resistance $R_{xx}$ for Ge(111) subsurface within regions I and II}
\label{LRG}
\begin{figure}[htbp]
\centering
\flushleft
\includegraphics[width=1.0\linewidth,clip]{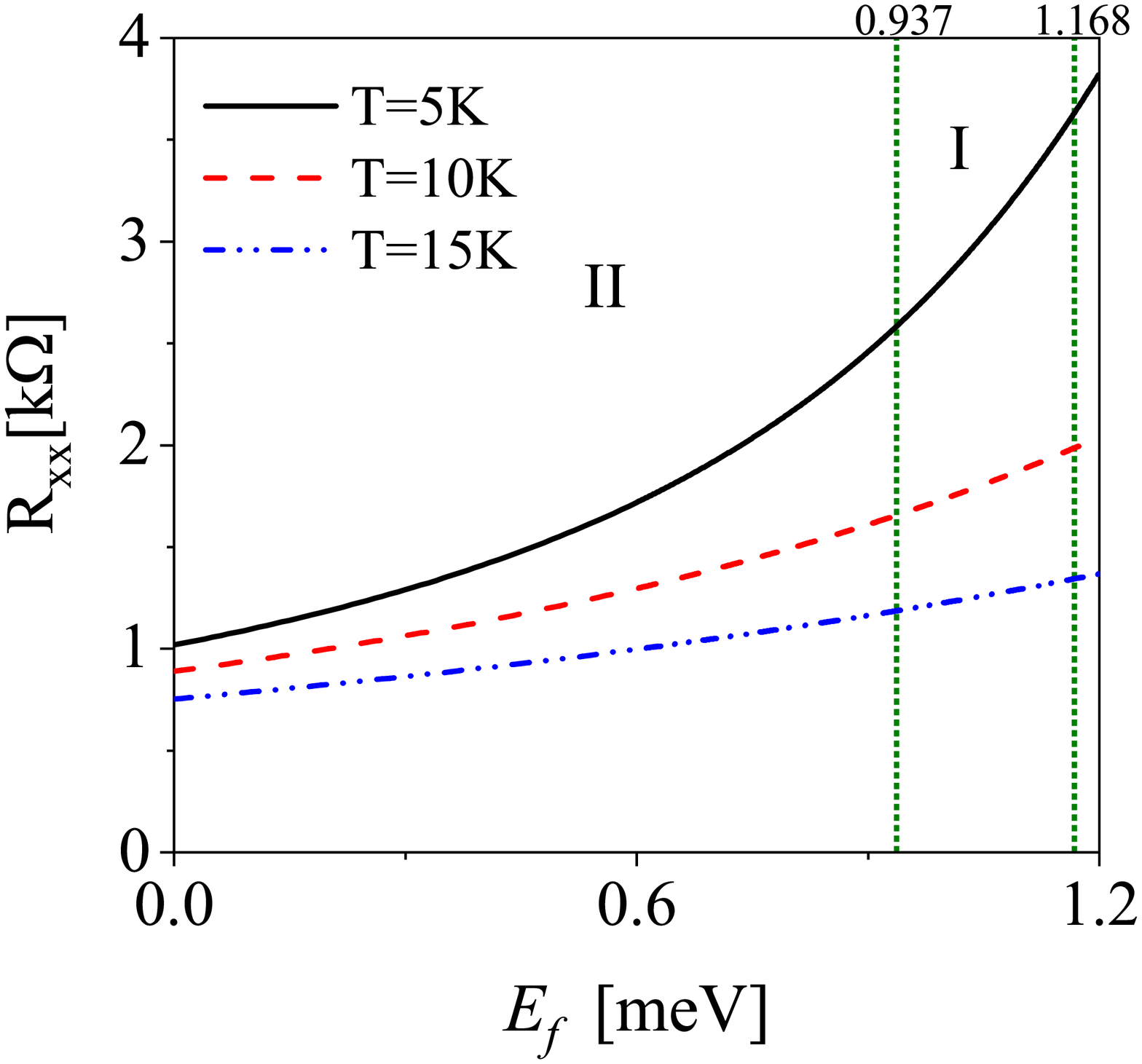}
\caption{The longitudinal Resistance $R_{xx}$ for Ge(111) subsurface as function of Fermi energy within the region I and II for different temperatures $T$. Parameters used: $\alpha_\text{R}=-0.2~eV{\AA}$, $g=2$, $\tau=2.88\times 10^{-11}s$, $l=120~\mu m$ and $w=30~\mu m$.}
\label{figure4}
\end{figure}

Since the largest values of $|\alpha_{xx}^{(2)}|$ appear in region I and II at low temperature, the longitudinal resistance $R_{xx}=l/(\sigma_{xx}w)$  will be calculated within regions I and II in this section. Through the standard Boltzmann equation within the relaxation time approximation, the longitudinal conductivity $\sigma_{xx}$ can be determined as \cite{Ziman}
\begin{equation}
\begin{aligned}
\sigma_{xx}&=-\frac{e^{2}}{\hbar}\frac{\tau}{(2\pi)^{2}}\int d\mathbf{k}v_{x}\frac{\partial f_{0}}{\partial k_{x}}\\
&=-\frac{e^{2}\tau}{(2\pi)^{2}}\int d\mathbf{k}(v_{x})^{2}\frac{\partial f_{0}}{\partial \epsilon_\mathbf{k}}.
\end{aligned}
\label{Appen-C-con}
\end{equation}
The energy $\epsilon_\mathbf{k}$ for Ge(111) subsurface is given in Eq.~(\ref{Enery}). Figure \ref{figure4} displays the resistance $R_{xx}$ of Ge subsurface as function of Fermi energy in region I and II for different temperatures. The length $l$ and width $w$ are taken as $120~\mu m$ and $30~\mu m$, respectively. As expected, $R_{xx}$ increase with increase of Fermi energy owing to the decrease of hole concentration.

\end{document}